\documentclass[english,usenatbib]{mn2e}

\usepackage{aas_macros}
\usepackage[english]{babel}
\usepackage[utf8x]{inputenc}
\usepackage{todonotes}
\usepackage{units}
\usepackage{amsmath}
\usepackage{graphicx}
\usepackage{url}
\usepackage[pdftitle={The BlueTides Simulation: First Galaxies and Reionization},pdfauthor={Y. Feng et al.}]{hyperref}
\title{The BlueTides Simulation: First Galaxies and Reionization}

\author[Yu Feng et. al]{Yu Feng\thanks{mailto:yfeng1@berkeley.edu}$^{1,2}$,
Tiziana Di-Matteo$^1$, Rupert A. Croft$^1$, 
Simeon Bird$^{1}$, 
Nicholas Battaglia$^{1,3}$
\and
Stephen Wilkins$^{4}$
\\
$^1$McWilliams Center for Cosmology, Carnegie Mellon University, Pittsburgh PA, 15213 \\
$^2$Berkeley Center for Cosmological Physics, University of California, Berkeley, Berkeley CA, 94720  \\
$^3$Department of Astrophysical Sciences, Princeton University, Princeton, NJ 08544 \\
Astronomy Center, Department of Physics and Astronomy, University
of Sussex, Brighton, BN19QH, UK \\
}

\newcommand{\kpch}{\,h^{-1}\unit{Kpc}}
\newcommand{\mpch}{\,h^{-1}\unit{Mpc}}
\newcommand{\msunh}{\,h^{-1}\unit{M_\odot}}
\newcommand{\software}[1]{{\small #1}}

\begin{document}
\maketitle

\begin{abstract}
We introduce the BlueTides simulation and report initial results for the luminosity functions of the first galaxies and AGN, and their contribution to reionization. 
BlueTides was run on the BlueWaters cluster at NCSA from $z=99$ to $z=8.0$ and includes 2$\times$7040$^3$ particles in a $400\,h^{-1}$Mpc per side box, making it the largest hydrodynamic simulation ever performed at high redshift. BlueTides includes a pressure-entropy formulation of smoothed particle hydrodynamics, gas cooling, star formation (including molecular hydrogen), black hole growth and models for stellar and AGN feedback processes.
The star formation rate density in the simulation is a good match to 
current observational data at $z\sim 8-10$. We find good agreement between observations and the predicted
galaxy luminosity function in the currently observable range $-18\le M_{\mathrm UV} \le -22.5$ with some dust extinction required to match the abundance of brighter objects. 
BlueTides implements a patchy reionization model that produces a fluctuating UV background. 
BlueTides predicts number counts for galaxies fainter than current observational limits which are consistent with extrapolating the faint end slope of the luminosity function with a power law index $\alpha \sim -1.8$ at $z\sim 8$ and redshift dependence of $\alpha \sim (1+z) ^{-0.4}$. The AGN population has a luminosity function well fit by a power law with a slope $\alpha \sim -2.4$ that compares favourably with the deepest CANDELS-Goods fields. 
We investigate how these luminosity functions affect the progress of reionization, and find that 
a high Lyman-$\alpha$ escape fraction ($f_\mathrm{esc} \sim 0.5$) is required if galaxies dominate the ionising photon budget during reionization.
Smaller galaxy escape fractions imply a large contribution from faint AGN (down to $M_\mathrm{UV} = -12$) which results in a rapid reionization, disfavoured by current observations.
\end{abstract}

\section{Introduction}
Recent deep observations using the Hubble Space Telescope have detected a plethora of objects at ever higher 
redshift \citep{2014arXiv1403.4295B, 2014arXiv1409.1228O, 2014arXiv1412.1472M} and 
measured the galaxy UV luminosity function at $z \leq 10$. 
At these redshifts microwave background measurements suggest 
that a substantial fraction of the Universe is still neutral \citep{2013ApJS..208...19H}, and these observations
may therefore probe the epoch of reionization. 
In the near future, next-generation space missions such 
as JWST \citep{JWST} and WFIRST \citep{WFIRST} will increase the number of available samples by several orders 
of magnitude. These are expected to detect the sources which produce 
the ionizing photons that drive reionization.

The formation of these objects is driven by non-linear gravitational collapse 
and so understanding them requires cosmological hydrodynamic simulations. The presence of an ionizing background may
affect galaxy formation \citep{MadauPaper}. To model the processes governing reionization it 
is desirable to simultaneously include scales of a few hundred Mpc, the characteristic size of ionization bubbles, 
and, to resolve the formation of galactic halos, scales of a few kpc. 

We present results from BlueTides, the largest 
cosmological hydrodynamic simulation yet performed, enclosing a box $400 \mpch$ on a side, with 
a smoothing length of $1.5 \kpch$, and including $2\times 7040^3$ particles-- a total of $0.7$ trillion particles.
Our high resolution allows us 
to study the formation of disc galaxies \citep{diskpaper}, while the large volume allows study of the progress of reionization.

We include a number of physically relevant processes, including star formation incorporating 
the effects of molecular hydrogen formation 
\citep{2011ApJ...729...36K}, energetic feedback from supernovae \citep{2010MNRAS.406..208O}, and feedback from super-massive black holes \citep{2005Natur.433..604D}.
We include for the first time in a simulation of this size a model 
for patchy reionization which varies the optical depth based on local density \citep{2013ApJ...776...81B}. 

Several large volume simulations have recently been performed. 
In particular, MassiveBlack I was the largest volume 
hydrodynamic simulation \cite{2012ApJ...745L..29D} to study reionization at $z > 4.75$.  
MassiveBlack II simulated a $100 \mpch$ volume at substantially improved resolution to $z=0$ \citep{2014arXiv1402.0888K}.
Illustris included a volume and resolution comparable to MassiveBlack II, 
but with improved prescriptions for the effect of energy injection from supernovae \citep{2010MNRAS.406..208O}.
The Eagle simulation is similar in size to MassiveBlack II and Illustris, 
but with a different approach to sub-grid modeling which allows improved agreement 
with observations by weakening requirements for numerical convergence with resolution \citep{2015MNRAS.446..521S}. 
Concurrently, dark matter only simulations have continued to increase in size, reaching loads of 
trillions of particles \cite{2014arXiv1410.2805H}.

BlueTides is based on the simulation code used in MassiveBlack I \& II, \software{P-Gadget3}~\citep{2005MNRAS.364.1105S,2012ApJ...745L..29D,2014arXiv1402.0888K}.
The simulation encloses a volume comparable to MassiveBlack I, has a resolution comparable to MassiveBlack II, 
and includes a stellar feedback model similar to that of Illustris. Our particle load is ten times larger than that of MassiveBlack I, 
which was previously the largest hydrodynamic simulation.

In Section \ref{sec:methods} we present our methods, explaining briefly the computational techniques necessary to perform a simulation of this magnitude.
In Sections \ref{sec:simstats} and \ref{sec:simstats2} we examine the basic statistics of objects within the simulation. We show the galaxy and AGN UV luminosity functions and star formation rates, and we present fits to these functions for easy comparison to future observations.
In Section \ref{sec:reion} we compute the sources of ionising photons within our model and use them to examine features of reionization. Finally we conclude in Section \ref{sec:conclusion}.

\section{BlueTides: Software and Sub-grid Physics}
\label{sec:methods}
\begin{figure*}
\includegraphics[width=\textwidth]{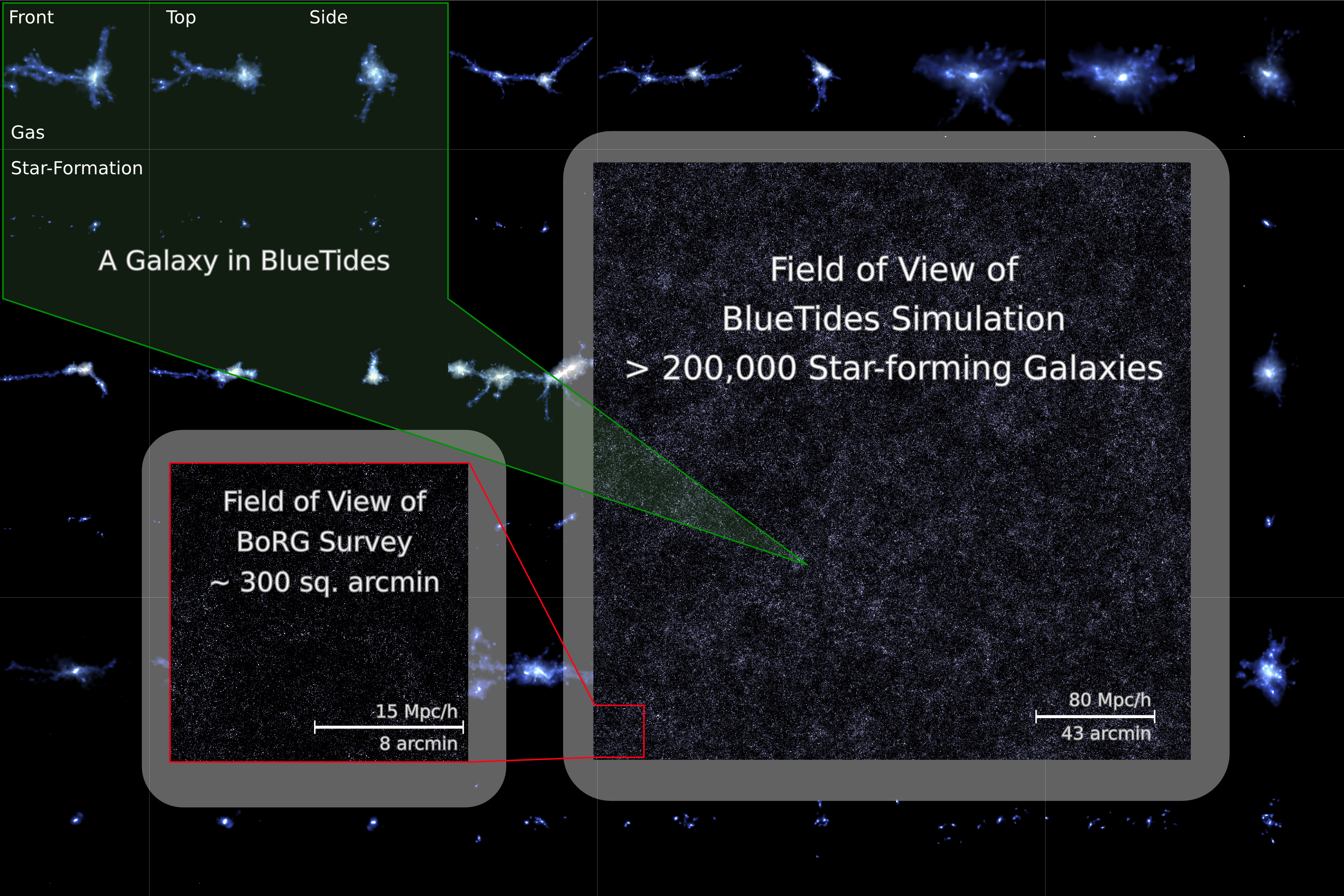}
\caption[Overview of the BlueTides Simulation]{Overview of the BlueTides Simulation. 
Left inset: a field of view from the BlueTides simulation at $z=8$ with the same total area as the BoRG survey \citep{2011ApJ...727L..39T}.
Right inset: the field of view of the entire BlueTides simulation at $z=8$.
Background: galaxies in BlueTides.
Even rows: top, front, and side views of the gas component of FoF halos.
Odd rows: top, front, and side views of the star formation rate surface density of FoF halos.
}
\label{fig:pretty-figure}
\end{figure*}

The BlueTides simulation was performed on the BlueWaters cluster at the National Center for Super-computing Applications (NCSA).
We operated the production run on a total of $648,000$ Cray XE compute core of BlueWaters.
This is the largest cosmological hydrodynamic simulation to date, containing a simulation volume roughly $300$ times larger than 
the largest observational survey at redshift $8-10$ \citep{2011ApJ...727L..39T}. 
This extraordinary size allows 
us to easily compare our results to current and future observations, 
and obtain a representative sample of the first galaxies 
which may have driven reionisation. A visual overview of the simulation is shown in Figure \ref{fig:pretty-figure}.

Halos in BlueTides are identified using a Friends-of-Friends algorithm with a linking length of $0.2$ times the mean particle separation \citep{1985ApJ...292..371D}.
We have not performed sub-halo or spherical over-density finder algorithms on BlueTides due to limits on the scalability of current implementations.
We will however investigate the mass function with spherical over-density finders in a follow up work.

\subsection{Computing: Improved performance at Peta-Scale}
At the particle numbers reached by our simulation, 
any unnecessary communication overhead can easily become a significant scalability bottleneck.
In order to allow the simulation to fully 
utilize the available computational power, we implemented several improvements to the speed and scalability of the code. 
Here we briefly list the substance of the most important changes, deferring a fuller description to \cite{YuFengInPrep}.

First, we substantially improved the scalability of the threaded tree implementation, which computes short-range particle interactions. This includes the gravitational force on small scales, hydrodynamic force, and the various feedback processes.
Our improved routine scales to $>30$ threads and at $8$ threads is twice as fast as the previous implementation in \software{P-Gadget3}.
Scalability improvements were achieved by eliminating OpenMP critical sections in favour of per-particle \software{POSIX} spin-locks, essentially making the thread execution wait-free at high probability.

Second, we replaced the default particle mesh gravity solver 
(used to compute long-range gravitational forces) based on FFTW
with a gravity solver based on a 2-d tile FFT library, \software{PFFT} \citep{doi:10.1137/120885887}. 
This allows a more efficient decomposition of particles to different processors, significantly improving the 
load and communication balance. To further simplify the book-keeping and reduce memory overhead, 
we switched to using Fourier-space finite differencing of gravity forces, as used by the N-body gravity solver HACC \citep{2014arXiv1410.2805H}. 
We also reduced the communication load by applying a sparse matrix compression of the local particle mesh.
In concert these changes removed the PM step as a scalability bottleneck. 

Third, at the high redshift covered by BlueTides, only a small fraction of the particles are in collapsed halos. 
Thus, to ease analysis, we produce two digest data sets on the fly in addition to the full simulation snapshots:
1) Particles-In-Group (PIG) files, which contain the attributes of all particles in the over-dense regions detected by the Friend of Friend groups;
2) Subsample files, containing a fair sub-sample of $1/1024$ of the dark matter and gas particles, and a full set of star and black hole particles.

Fourth, we implemented a histogram based sorting routine to replace the merge sort routine originally present in \software{P-Gadget3}.
Histogram based sorting routines have been shown to perform substantially better at scale \citep{edgerandvivek}, a result confirmed by our experience in BlueTides \citep{mpsort}.
Particles must be sorted when constructing the FoF group catalogues, and our new sort routine sped up FoF catalogue generation times by a factor of $10$.  
The source code of this sorting routine, \software{MP-Sort}, is available from \url{http://github.com/rainwoodman/MP-sort} to facilitate independently reuse.

Finally, we implemented a new snapshot format (\software{BigFile}) that supports transparent file-level striping and 
substantially eases post-production data analysis compared to multiple plain HDF5 files. Until recently, file-level striping (bypassing MPIO) has been the only way to achieve the full IO capability of Lustre file system for problems at our scale. We release the library for accessing the BlueTides simulation at \url{http://github.com/rainwoodman/bigfile}, together with Python language bindings for post-simulation data analysis. 

\subsection{Physics: Hydrodynamics and sub-grid modelling}
\label{sec:physics}
BlueTides uses the hydrodynamics implementation described in \cite{2014MNRAS.440.1865F}. 
We adopt the pressure-entropy formulation of smoothed particle hydrodynamics (pSPH) to solve the Euler equations \citep{2013MNRAS.428.2840H,2010MNRAS.405.1513R}.
The density estimator uses a quintic density kernel to reduce noise in SPH density and gradient estimation \citep{2012JCoPh.231..759P}.

\begin{table*}
  \begin{tabular}{ccl}
  \hline
    Name & Value & Notes \\
    \hline
    $\Omega_\Lambda$  & 0.7186  \\
    $\Omega_\mathrm{Matter}$ & 0.2814 &     Baryons + Dark Matter\\
    $\Omega_\mathrm{Baryon}$ &    0.0464  \\
    $h$  &  0.697 & Hubble parameter in units of $100\,\unit{km/s/\mathrm{Mpc}}$ \\
    $\sigma_8$  &      0.820  \\
    $n_s$  &     0.971\\
    $L_\mathrm{Box} $ &     $400\mpch$ & Length of one side of the simulation box. \\
    $N_\mathrm{Particle}$ &     $2\times7040^3$  &   Total number of gas and dark matter particles in the initial conditions \\
    $M_\mathrm{DM}$  &     $1.2\times10^7 h^{-1} \unit{M_\odot}$  &   Mass of a dark matter particle. \\
    $M_\mathrm{GAS}$  &    $2.36\times10^6 h^{-1} \unit{M_\odot}$ &     Mass of a gas particle in the initial conditions. \\
    $\eta_h$ &     1.0 &     SPH smoothing length in units of the local particle separation$^\dagger$. \\
    $\varepsilon_\mathrm{grav}$ &    $1.5\kpch$ &     Gravitational softening length. \\
    $N_\mathrm{Generation}$ &     4 &     Mass of a star particle as a fraction of the initial mass of a gas particle. \\
    $\mathrm{egy}_w / \mathrm{egy}_0$  &      1.0  &      Fraction of supernova energy deposited as feedback. \\
    $\kappa_w$  &     3.7 &    Wind speed as a factor of the local dark matter velocity dispersion. \\
    $E_\mathrm{SNII,51}$ &     1.0 &    Supernova energy in units of $10^{51}\,\unit{erg/s}$ \\
    $M^{(0)}_\mathrm{BH}$ &    $5\times10^{5}\msunh$ &    Seed mass of black holes.\\
    $M^{(0)}_\mathrm{HALO}$ &    $5\times10^{12}\msunh$ &    Minimum halo mass considered in black hole seeding.\\
    $\eta_\mathrm{BH}$ &    0.05 &    Black hole feedback efficiency. \\
    \hline
  \end{tabular}
  
  $^\dagger$ A value of 1.0 translates to 113 neighbour particles with the quintic kernel used in BlueTides.
  \caption{Parameters of the BlueTides Simulation}
  \label{tab:simparam}
\end{table*}

Table \ref{tab:simparam} lists the basic parameters of the simulation. Initial conditions are generated at $z=99$ using an initial power spectrum from \software{CAMB} \citep{Lewis:2002ah}.
Star formation is implemented based on the multi-phase star formation model in \cite{2003MNRAS.339..289S}, and incorporating several effects following \cite{2013MNRAS.436.3031V}.
Gas is allowed to cool both radiatively following \cite{1996ApJS..105...19K} and via metal cooling. We approximate the metal cooling rate by scaling a solar metallicity template according to the metallicity of gas particles, following \cite{2014MNRAS.444.1518V}.
We model the formation of molecular hydrogen, and its effect on star formation at low metallicities, 
according to the prescription by \cite{2011ApJ...729...36K}. 
We self-consistently estimate the fraction of molecular hydrogen gas 
from the baryon column density, which in turn couples the density gradient 
into the star-formation rate. 

A SNII wind feedback model \citep{2010MNRAS.406..208O} is included, which assumes wind speeds proportional to the local one dimensional dark matter velocity dispersion $\sigma_\mathrm{DM}$:
\begin{equation}
v_w = \kappa_w \sigma_\mathrm{DM} \,,
\end{equation}
where $v_w$ is the wind speed. $\kappa_w$ is a dimensionless parameter, which we take to be $3.7$ following \cite{2013MNRAS.436.3031V}.

We model feedback from active galactic nuclei (AGN) in the same way as in the MassiveBlack I \& II simulations, using the super-massive black hole model developed in \cite{2005Natur.433..604D}. 
Super-massive black holes are seeded with an initial mass of $5\times10^{5}\msunh$ in halos more massive than $5\times10^{10}\,h^{-1}\unit{M_\odot}$, while 
their feedback energy is deposited in a sphere of twice the radius of the SPH smoothing kernel of the black hole.

\begin{figure}
\includegraphics[width=\columnwidth]{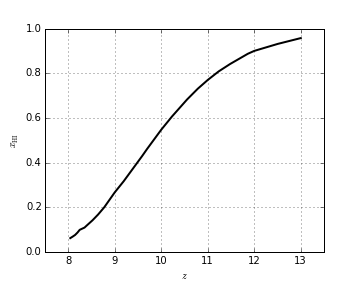}
\caption{Global neutral hydrogen fraction as a function of redshift in BlueTides.}
\label{fig:xhi}
\end{figure}

The large volume of BlueTides allowed us to include some of the effects of ``patchy`` reionization, where the amplitude of the ultraviolet background is spatially variable.
We model patchy reionization using a semi-analytic method based on hydrodynamic simulations performed with radiative transfer \citep[for more details see][]{2013ApJ...776...81B}. 
This method uses an evolved density field calculated from the initial conditions using second order Lagrangian perturbation theory to predict 
the redshift at which a given spatial region will reionize. 
In our fiducial reionization model, we set the mean reionization redshift at $z \sim 10$ based on the measurement of the optical depth, $\tau$, from the WMAP 9 year data release
\citep{2013ApJS..208...19H}. In regions that have been reionized, we assume the UV background estimated by \cite{2009ApJ...703.1416F}. The global neutral fraction in BlueTides evolves smoothly as a function of redshift, as seen in Figure \ref{fig:xhi}.

\section{Star Formation Rate}
\label{sec:simstats}

\begin{figure*}
\includegraphics[width=\textwidth]{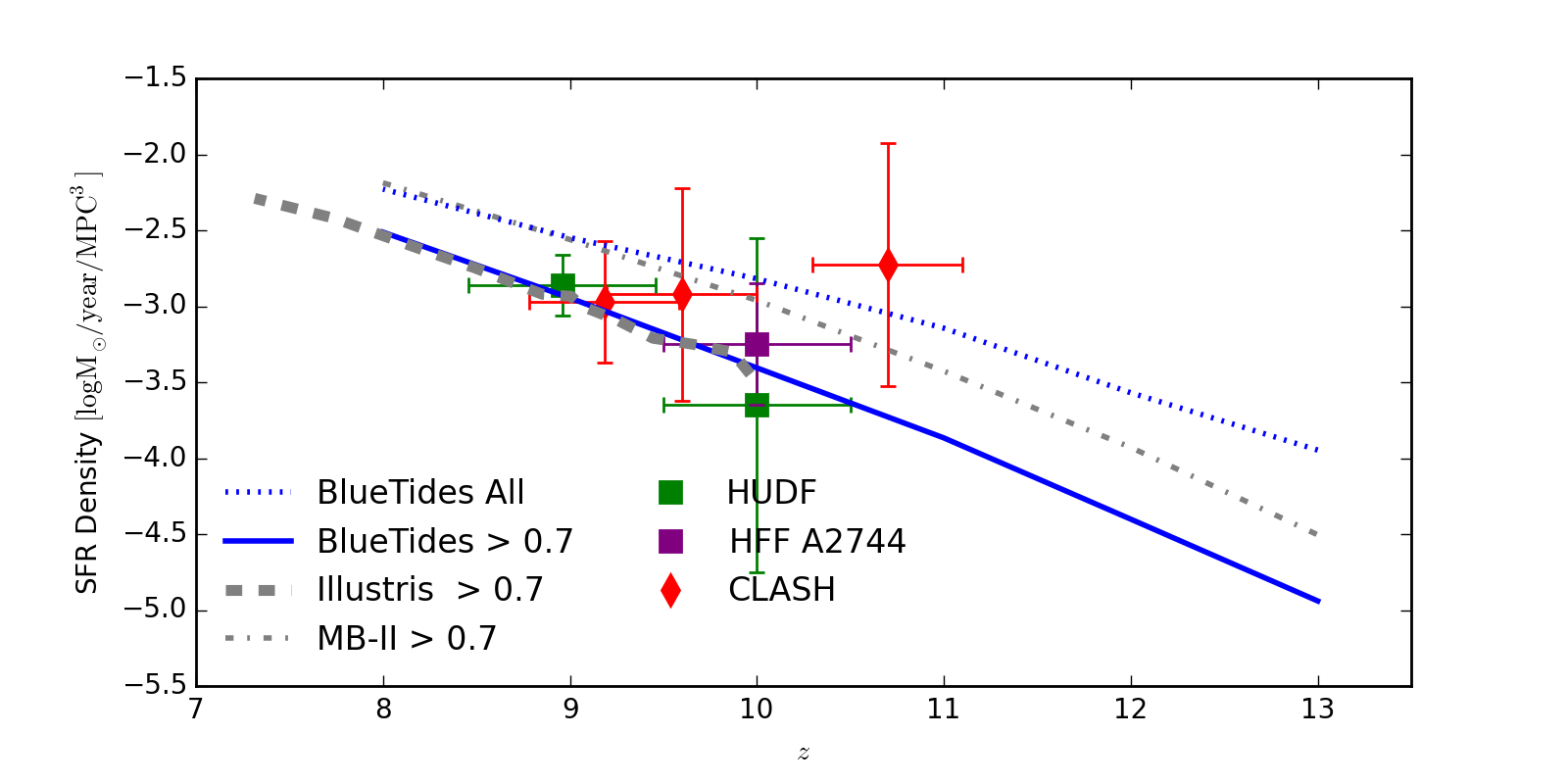}
\caption[Star-formation rate density in BlueTides]{The Star Formation Rate Density in the BlueTides simulation. 
Solid blue: star formation rate density of halos with star formation rate greater than 
$0.7\, \unit{M_\odot/year}$ from BlueTides.
Dashed black: star formation rate density of equivalent halos from Illustris
\citep{2014MNRAS.445..175G,2014MNRAS.444.1518V}.
Solid black: star formation rate density of equivalent halos from MassiveBlack II \cite{2014arXiv1402.0888K}.
Dotted blue: star formation rate density of all halos from BlueTides.
Dotted black: star formation rate density of all halos from MassvieBlack II.
Squares: estimates from HUDF \citep{2014ApJ...786..108O}, and HFF A2744\citep{2014arXiv1409.1228O}.
Diamonds: observational estimates from CLASH \citep{2014ApJ...795..126B,2012Natur.489..406Z,2013ApJ...762...32C}.
Illustris, CLASH, and HUDF lines are reproduced from \cite{2014arXiv1409.1228O}.
}
\label{fig:SFR-survey}
\end{figure*}
Figure \ref{fig:SFR-survey} shows the global star formation rate in BlueTides, together with observational constraints and several other simulations.
We show the total star formation density in the whole volume (dotted lines) and
the total star formation rate for halos with SFR $> 0.7\, \unit{M_\odot/year}$ (solid line). 
The latter is directly comparable with current observations and corresponds to the current observational limit of $M_\mathrm{UV} = -18$.
The SFR density in BlueTides smoothly increases with decreasing redshift.  


Several Hubble ultra deep field surveys have given estimates on the star formation rate density due to halos with $M_\mathrm{UV} < -18$. 
BlueTides halos at the observational limit typically contain a few thousand particles, and are thus well resolved.
The BlueTides predictions for the star formation rate is in good agreement with current observations, with the caveat that at these high 
redshifts observational uncertainty remains high.


Figure \ref{fig:SFR-survey} also compares the SFR density in BlueTides to that from two other recent (albeit smaller volume) simulations:
Illustris \citep{2014MNRAS.445..175G,2014MNRAS.444.1518V}, and MassiveBlack II \citep{2014arXiv1402.0888K}.
Illustris uses similar prescriptions for sub-grid feedback, but a different solver for the Euler equations, while MassiveBlack II uses 
substantially different sub-grid modelling, which is less effective at suppressing star formation in faint objects.
Neither of the other simulations include patchy reionization. 
The star formation rates for BlueTides and Illustris agree very well, while that for MassiveBlack II is a factor of a few larger. 
This suggests that the sub-grid feedback model dominates in controlling the star formation rate over both the choice of hydrodynamic method and the 
effect of reionization. It is however promising to see, given the differences in the simulations, that the
differences are within at most a factor of a few in SFR density. 

\begin{figure}
\includegraphics[width=\columnwidth]{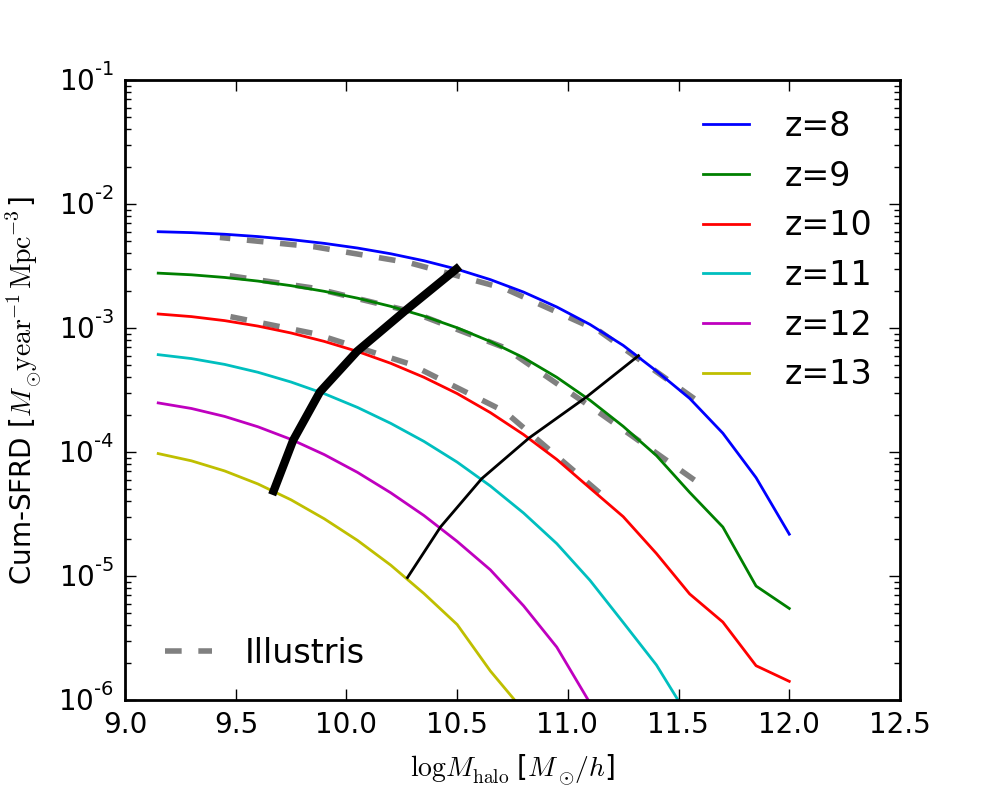}
\caption{Cumulative star-formation rate density in halos. 
Coloured solid: cumulative SFRD in BlueTides.
Gray dashed: cumulative SFRD in Illustris at $z=8, 9, 10$, from the Illustris public data release \protect\citep{2015arXiv150400362N}.
Thick black: contour of 50\% cumulative SFRD.
Thin black: contour of 10\% cumulative SFRD.
}
\label{fig:cumsfr}
\end{figure}

In Figure \ref{fig:cumsfr}, we show the cumulative star-formation rate density from galaxies of different halo mass. For comparison, we also calculated the cumulative SFR density of Illustris at three corresponding redshifts ($z=8, 9, 10$). 
We see that the mass threshold for 50\% of star-formation increases from $\sim 5\times 10^9 h^{-1} M_\odot$ at $z=13$ to $\sim 4\times 10^{10} h^{-1} M_\odot$ at $z=8$. Thus the contribution of small halos to the ionizing photon budget becomes increasingly important at higher redshift.
The larger volume in BlueTides means that it includes halos $10$ times more massive than the most massive halos in Illustris. These objects contribute less than 10\% of the total star-formation density at $z=8, 9$, and $10$.

\section{Stellar and AGN UV Luminosity Functions}
\label{sec:simstats2}

In this Section, we report the stellar AGN luminosity functions in BlueTides. We first describe  our source detection method, using Source Extractor to validate
the results of an FoF halo finder at these high redshifts (Section \ref{sec:obscomp}). 
We then describe the stellar UV luminosity function (Section
\ref{sec:stellarlumin}),
the dust attenuation model (Section \ref{sec:dust}), and the faint-end slope of the luminosity function (Section \ref{sec:slope}).
We assemble the stellar luminosity function from BlueTides and compare to observations and other simulations (Section \ref{sec:stellarobs}). Finally we report  the AGN luminosity function(Section \ref{sec:AGN}).

\subsection{The identification of Galaxies in BlueTides: Source Extractor vs FoF}
\label{sec:obscomp}

The Friend-of-Friends (FoF) algorithm considers only the spatial positions of particles and 
can sometimes artificially group dynamically distinct objects into one halo. 
In this section we compare the luminosity function estimated from FoF catalogues to that which would be estimated by performing standard observational techniques
and show that the difference is small.

Simulations and observations have long been defining objects in different ways. 
Even the sub-halos in simulations do not directly translate to any imaging survey catalogs. 
In imaging surveys, 
objects are identified by selecting peaks in a 2-dimensional image, 
and the total luminosity depends on an aperture radius. 
\citep[we refer the readers to][and references herein]{2014MNRAS.445..239S} 
The canonical implementation of such an algorithm is Source Extractor \citep{1996A&AS..117..393B};
We use SEP, a reimplementation of source extractor into python \citep{SEP}.

We produce a mock $2-d$ survey image by projecting the BlueTides simulation box along one axis. 
The star formation rates of particles are distributed into pixels using a Gaussian kernel scaled by their SPH smoothing lengths with \software{GAEPSI} \citep{2011ApJS..197...18F}.  
We do not attempt to model instrumental noise in the mock image.
The final star formation surface density image has $(2\times 10^5)^2$ pixels, 
with a spatial resolution of $2\kpch$ per pixel ($\sim 0.06 \unit{arcsec}$). 
This image is then divided into $100$ non-overlapping equal sized sub-volumes $x-y$ plane. 
Each sub-volume has a volume of $40\times40\times400(\mpch)^3$.
This allows us to estimate the cosmic variance in e.g. the luminosity functions in fields comparable to those observed. 
We note that $400\mpch$ roughly corresponds to a redshift width of $\Delta z\sim 1$ at $z=8.0$, and hence each image chunk roughly corresponds to the full volume of the BoRG survey.

We run SEP on each of the image chunks, afterwards combining them to assemble the full catalogue. 
Edge effects can be safely neglected because the area of the images are much larger than the area of the edges.
We use a near-zero threshold ($1\times10^{-6}\,\msunh\kpch^2$) to include all pixels that have non-zero star formation surface density. 
The integrated star formation rate of the objects is measured with an aperture size of $8.7$ pixels ($0.5\,\unit{arcsec}$), 
which roughly corresponds to the aperture used by the BoRG survey \citep{2011ApJ...727L..39T}.  
Figure~\ref{fig:sepatwork} shows visually the process of identifying galaxies using source extractor.
We compute UV magnitudes from the star-formation rate of the SEP catalogue and the FoF catalogue using \citep{2011MNRAS.414.1927S}:

\begin{figure*}
\includegraphics[height=1.0\textwidth]{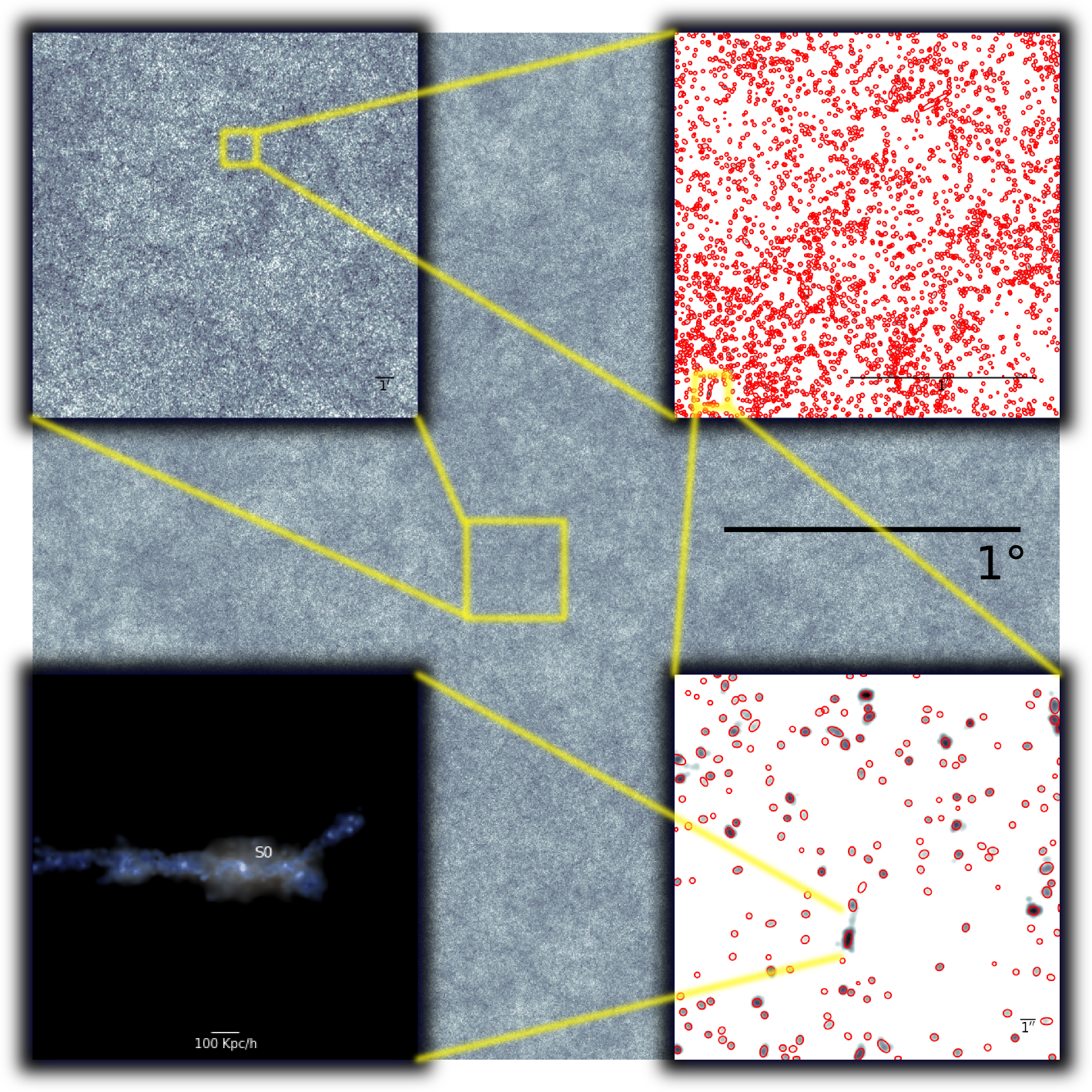}
\caption{Detecting objects from the mock star-formation intensity image.
Background: the star-formation intensity image of the full projection of BlueTides ($400\mpch$ per side). We note that the image is strikingly uniform because of the thickness of the projection.
Top left: the star-formation intensity image of a single chunk, $40\mpch$ per side.
Top right: all objects identified in a field of view with a 10 times zoom, $4\mpch$ per side. SEP objects are marked in red.
Bottom right: a further zoomed-in view of the top right panel, to show the identified objects more clearly.
}
\label{fig:sepatwork}
\end{figure*}

\begin{equation}
M_\mathrm{UV} = - 2.5 \log_{10}(\Psi)  - 18.45 \,.
\label{eq:SFTOUVGAL}
\end{equation}
We exclude faint objects with $M_\mathrm{UV} > -14$, corresponding to unresolved halos typically containing less than $50$ dark matter particles, and for the moment neglect dust extinction, which we will discuss in Section \ref{sec:dust}.

\begin{figure}
\includegraphics[width=\columnwidth]{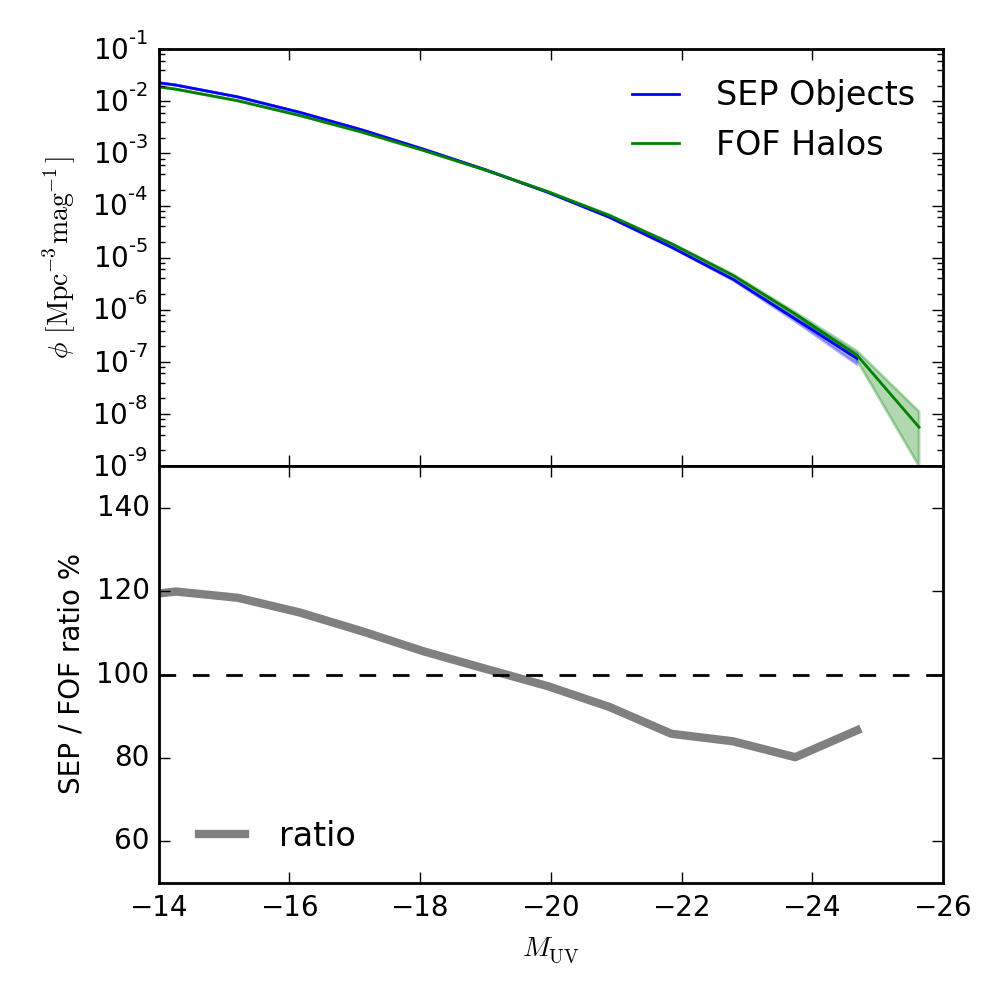}
\caption{UV luminosity function of galaxies found with source extractor compared to that of FoF halos at $z=8$. Green: SEP catalogue. Blue: FoF catalogue. Solid grey: the ratio between SEP and FoF (axis on right).
}
\label{fig:fofsep}
\end{figure}

As shown in Figure~\ref{fig:fofsep}, we find that the luminosity functions constructed from the FoF halo catalogue and the SEP imaging catalogue differ by less than $20$\%. The main differences are that the SEP luminosity function includes fewer bright objects than the FoF luminosity function and more faint objects. Quantitatively, SEP is $\sim 80\%$ of FoF at $M_\mathrm{UV} > -22$ and $\sim 120\%$ of FoF at $M_\mathrm{UV} < -18$. These differences can be understood by noting that bright (massive) FoF halos are often a cluster of smaller objects which SEP tends to identify as separate galaxies. The fixed aperture in SEP however tends to enhance the UV of smaller objects. These effects are smaller at these high redshifts, where large groups have not yet formed.
Because these differences are small, we will use the FoF catalogue for the rest of this work. 

\subsection{Stellar Luminosity Functions}

\subsubsection{Intrinsic Stellar Luminosity Functions}
\label{sec:stellarlumin}
\begin{figure}
\includegraphics[width=\columnwidth]{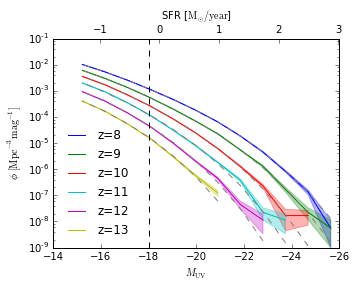}
\caption{Evolution of the intrinsic UV Luminosity function
  with redshift from $z=8-13$ (colours online).
  Shaded regions show the $1-\sigma$ sample variance of the mass functions. Dashed lines: best fit modified Schechter model.
  The vertical dash line corresponds to the current observation limit of $M_\mathrm{UV} \sim -18$.
}
\label{fig:FoFUV}
\end{figure}
Figure \ref{fig:FoFUV} shows the intrinsic galaxy UV luminosity function
evolving from $z=13$ to $z=8$. The luminosity function is computed from all galaxies within the simulation. Shaded areas represent $1-\sigma$ uncertainty of the mass function estimated from 100 sub-volumes. The luminosity function's evolution in redshift is largely described by an increase in amplitude, with minor evolution of the faint-end slope. Figure~\ref{fig:FoFUV} also shows a fit of the intrinsic UV luminosity function to a modified Schechter model, which captures the simulated luminosity function well at all redshifts. The modified Schechter model is

\begin{equation}
\begin{aligned}
\ln \phi(M) &= \ln \phi^\star + \ln A \\
    &+ A (M^\star -M)(1 - \alpha_L) \\
    & - 10^{0.1 (M^\star -M)},
\end{aligned}
\label{eq:schechter-replaced}
\end{equation}

This model differs from the usual Schechter model (see equation \ref{eq:schechter} below) only in that the coefficient that determines the rate of extinction of bright end galaxies is changed from $0.4$ to $0.1$. The best fit values are shown in Table~\ref{tab:galfit}. The modified model agrees with the luminosity function at the $5\%$ level for the whole luminosity range of the simulation.
Note that the change in the bright end coefficient means that one should not directly compare these fit parameters with those obtained from a Schechter model. We also do not use this model to extrapolate the luminosity function (although with a minor $10\%$ difference in the photon budget, using it would not change our conclusions).

\begin{table}
  \caption{Best-fit parameters of the modified Schechter model for the galaxy UV luminosity functions at $z=8 - 13$. Parameters are fit using Equation \ref{eq:schechter-replaced}.}
\label{tab:galfit}
\centering
\begin{tabular}{cccc}
\hline
$z$ & $\alpha_L$ & $\log \phi^\star$ & $M_\mathrm{UV}^\star$ \\
\hline
8 & $-1.54\pm0.01$ & $-4.04\pm0.07$ & $-15.99\pm0.09$ \\
9 & $-1.59\pm0.02$ & $-4.17\pm0.14$ & $-15.44\pm0.19$ \\
10 & $-1.55\pm0.04$ & $-3.66\pm0.18$ & $-14.09\pm0.28$ \\
11 & $-1.51\pm0.07$ & $-3.42\pm0.21$ & $-13.03\pm0.40$ \\
12 & $-1.40\pm0.07$ & $-3.42\pm0.10$ & $-11.78\pm0.37$ \\
13 & $-1.32\pm0.10$ & $-3.82\pm0.08$ & $-10.98\pm0.44$ \\
\hline
\end{tabular}
\end{table}

\subsubsection{Dust Extinction}
\label{sec:dust} 
There is evidence that the highest luminosity early galaxies are significantly dust obscured \citep{wilkins13,cen14}. 
To produce luminosity functions more comparable to observations, we adopt the screening model from \cite{joung09}. This attenuates the UV luminosity from a pixel in the face-on image of a galaxy by a fraction, $f_{\rm UV}$, proportional to the metal-mass density in that pixel.
The value of $f_{\rm UV}$ determines the extinction coefficient $A_{\rm UV}$. 
We apply this dust model to the bright individual galaxies at $z=8$, and find that the dust extinction in UV band is fitted by 
\begin{equation}
M^d_\mathrm{UV} - M^i_\mathrm{UV} =
\exp \left[
-\frac
{M^i_\mathrm{UV} + 22.61}
{1.72}\right]
\label{eq:dustmodel}
\end{equation}
where $M^i_\mathrm{UV}$ is the intrinsic UV luminosity and $M^d_\mathrm{UV}$ is the UV luminosity with a dust correction. Equation \ref{eq:dustmodel} produces dust extinction of $A_{\rm UV} \sim 1$ for $M_{\rm UV}=-21$ galaxies, which agrees with the upper limit inferred from the UV slope of high redshift galaxies by \cite{wilkins13} at $z=8.0$.

\begin{figure}
\includegraphics[width=\columnwidth]{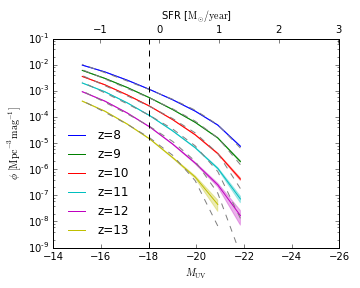}
\caption{Evolution of the UV Luminosity function with dust extinction
  with redshift $z=8, 9, 10, 11, 12, 13$. (colours online)
  Shaded regions show the $1-\sigma$ uncertainty of the mass functions. Dashed lines: best fit modified Schechter model (see Equation \ref{eq:schechter})
  The vertical dashed line corresponds to the current observational detection limit of $M_\mathrm{UV} \sim -18$.
}
\label{fig:FoFUV-DUST}
\end{figure}

Figure \ref{fig:FoFUV-DUST} shows the galaxy UV luminosity function with dust extinction at $1500$\AA, from $z=13$ down to $z=8$. We also show the results of a Schechter fit \citep{1976ApJ...203..297S} to the observed luminosity functions. The Schechter model is widely used to parametrise luminosity functions.
We use the form provided by \cite{2012MNRAS.420.1606J}.
\begin{equation}
\begin{aligned}
\ln \phi(M) &= \ln \phi^\star + \ln A \\
    &+ A (M^\star -M)(1 - \alpha_L) \\
    & - 10^{0.4 (M^\star -M)},
\end{aligned}
\label{eq:schechter}
\end{equation}
where $A = 0.4 \ln 10$. Parameters are estimated using $\chi^2$ fitting over $\ln \phi$, assuming uncorrelated errors (estimated from the sub-volumes). 
The best fit parameters are reported in Table \ref{tab:galfit-s}. Note that the Schechter model does not describe the BlueTides luminosity functions at high redshift ($z>10$), and systematically under-fits the bright end luminosity function, even after including dust extinction.

\begin{table}
  \caption{Best-fit Schechter Model parameters at $z=8 - 13$ for galaxy stellar UV luminosity functions including dust extinction. Parameters are as described in Equation \ref{eq:schechter}.}
\label{tab:galfit-s}
\centering
\begin{tabular}{cccc}
\hline
$z$ & $\alpha_L$ & $\log \phi^\star$ & $M_\mathrm{UV}^\star$ \\
\hline
8 & $-1.84\pm0.03$ & $-8.90\pm0.21$ & $-20.95\pm0.15$ \\
9 & $-1.94\pm0.03$ & $-9.74\pm0.24$ & $-20.77\pm0.17$ \\
10 & $-2.01\pm0.04$ & $-10.27\pm0.33$ & $-20.39\pm0.22$ \\
11 & $-2.07\pm0.05$ & $-10.77\pm0.42$ & $-20.00\pm0.27$ \\
12 & $-2.12\pm0.06$ & $-11.40\pm0.50$ & $-19.65\pm0.31$ \\
13 & $-2.13\pm0.06$ & $-11.71\pm0.49$ & $-19.11\pm0.30$ \\
\hline
\end{tabular}
\end{table}
\subsubsection{Faint-end Slope}
\label{sec:slope}
\begin{figure}
\includegraphics[width=\columnwidth]{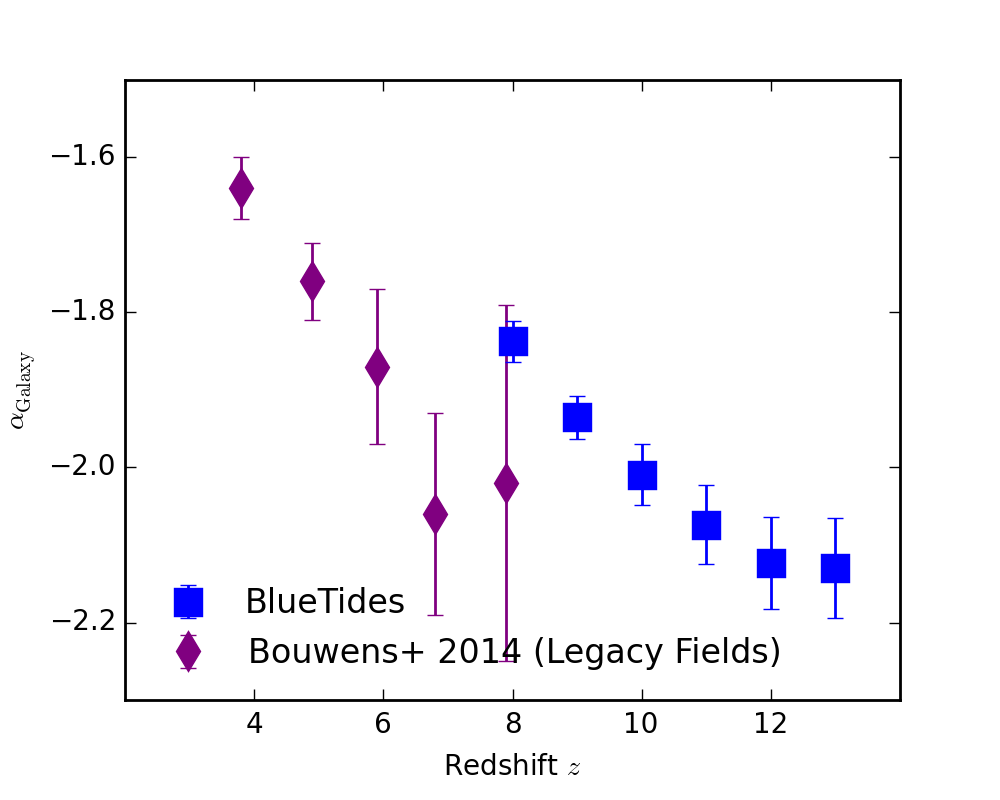}
\caption[Evolution of the faint-end slope of the galaxy UV luminosity function]{Evolution of the faint-end slope of the galaxy UV luminosity function. The purple diamonds are the observed slope from Hubble legacy surveys \protect\citep{2014arXiv1403.4295B}. Blue circles show the slope from BlueTides.}
\label{fig:galf-slope-ev}
\end{figure}
We show the redshift evolution of the faint-end slope of the dust extincted galaxy UV luminosity function in BlueTides in Figure \ref{fig:galf-slope-ev}. The best fit model for the evolution is
\begin{equation}
\alpha_\mathrm{Galaxy}(z) = -0.756 (1+z)^{0.41}
\end{equation}

The slope of the faint end of the UV luminosity is consistent with that inferred by \cite{2014arXiv1403.4295B} and its evolution with redshift implies moderate steepening, again consistent with an extrapolation of the observed slope evolution up to $z=12$, which includes an evolution of $M/L$ ratio $\propto (1+z)^{-1.5}$ due to the evolution of the Halo Mass Function\footnote{in Fig~\ref{fig:galf-slope-ev} we ignored the $z=10$ estimate from \cite{2014arXiv1403.4295B}, as the authors manually set the faint-end slope at $z=10$}.
The faint end ($M_\mathrm{UV} > -20$) of the UV luminosity function in BlueTides is barely affected by the dust extinction model, thus the redshift-slope relation we give here is suitable for inputs of reionization calculations.

\subsubsection{Comparison with Observations and Other Models}
\label{sec:stellarobs}

\begin{figure*}
\includegraphics[width=\textwidth]{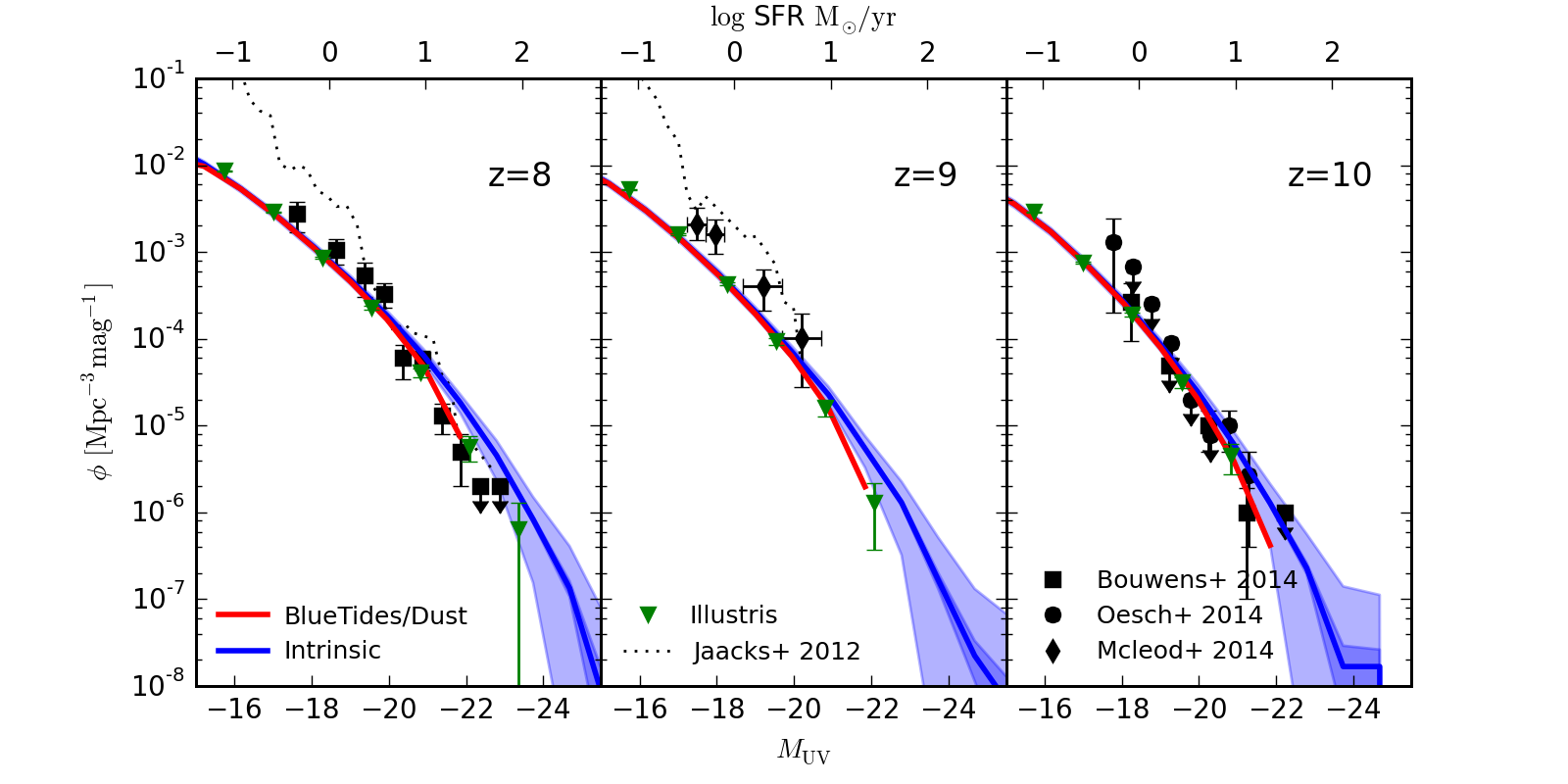} 
\caption[Luminosity function comparison]{Comparing UV Luminosity functions with observations.
Solid blue: Intrinsic UV luminosity function in BlueTides. The coloured bands shows the cosmic variance in a survey volume of the size of the BoRG survey \citep{2014arXiv1403.4295B}.
Solid red: Dust-reddened UV luminosity function in BlueTides.
Dotted black: Luminosity functions from the simulations of \cite{2012MNRAS.420.1606J}. Note that they do not provide a simulated luminosity function at $z=10$; also the largest volume at $z=9$ was a $34 \mpch$ box.
Green wedges: intrinsic UV luminosity function from the Illustris public data release \cite{2015arXiv150400362N}.
Black dashed line: intrinsic UV luminosity function from the MassiveBlack-II public data  \cite{2014arXiv1402.0888K}.
Black squares: Observed luminosity functions at $z=8$ and $z=10$ from \cite{2014arXiv1403.4295B}.
Black diamonds: Observed luminosity function at $z=9$ from \cite{2013MNRAS.432.2696M,2014arXiv1412.1472M}.
Black circles: Observed luminosity function from four bright galaxies at $z\sim 10$ from \cite{2014ApJ...786..108O}.
}
\label{fig:tension}
\end{figure*}

\cite{2014arXiv1403.4295B} assembled and reanalyzed the UV luminosity function evolution from $z=10$ to $z=4$ based on all currently available 
legacy Hubble surveys.
The total cumulative area is close to $1000\,\unit{arcmin^2}$, spread over a wide redshift range \citep[see also][]{2013ApJ...763L...7E, 
2014arXiv1410.5439F, 2010ApJ...714L.202T}. The most recent measurements at $z=10$ were published in  \cite{2014ApJ...786..108O}. 
Figure \ref{fig:tension} compares a compilation of these observational data to stellar UV luminosity functions from BlueTides.
We show the BlueTides luminosity functions extracted from subfields with size roughly the area of the BoRG survey \citep{2014arXiv1403.4295B}, the legacy field with the 
currently largest area. As this is a smaller volume than the full simulation, we can use the differences between sub-volumes to estimate sample variance from current observations, 
which is shown by the shaded areas in Figure~\ref{fig:tension}. The intrinsic luminosity produces more bright galaxies than are observed, a discrepancy which is marginally significant compared to cosmic variance. After applying a dust extinction correction, this slight tension with observations largely disappears, suggesting that dust corrections are indeed significant for the brightest galaxies, even at these high redshifts. As we shall discuss in Section~\ref{sec:AGN}, another possibility is that the brightest sources
host a significant AGN which may make their detection in the galaxy samples harder.

We show comparisons with two other recent high redshift simulations with a maximal volume of $100 \mpch$. \cite{2012MNRAS.420.1606J} performed several simulations at different resolutions to investigate the shape and slope of high redshift galaxy UV luminosity function. The luminosity functions of \cite{2012MNRAS.420.1606J} have faint end slopes which are steep compared to observations, producing substantially too many stars in small objects. This is likely due to their stellar feedback \citep{2011MNRAS.410.2579C} being insufficiently effective at suppressing star formation. We also show the corresponding UV luminosity function at $z=8, 9, 10$ from the public data of Illustris \citep{2015arXiv150400362N}. Due to the larger volume in BlueTides, 
Illustris produces fewer bright objects and cannot be compared to BlueTides at the most massive end. However, at the faint end of the luminosity function ($M_\mathrm{UV} < -20$), BlueTides and Illustris agree at the $10$\% level. This suggests that a stellar feedback model which very efficiently suppresses star formation in small halos is the most important ingredient when matching the luminosity function at high redshift, just as at low redshift.

\subsection{AGN Luminosity Function}
\label{sec:AGN}
As described in Section \ref{sec:methods}, the BlueTides simulation models AGN via a self-regulated super-massive black hole model following \cite{2005Natur.433..604D}. 
Given a mass accretion rate $\frac{dM_\mathrm{BH}}{dt}$ the bolometric luminosity of AGN is 
\begin{equation}
	L = \eta \frac{dM_{\mathrm{BH}} c^2}{dt},
\end{equation}
where $\eta=0.1$ is the mass-to-light conversion efficiency in an accretion disk.

We convert the bolometric luminosity of AGN in the simulation to a UV magnitude using \citep{2012MNRAS.425.1413F}:
\begin{equation}
	M_\mathrm{UV} = -2.5 \log_{10} \frac{L_\mathrm{BOL}}{f_B \nu_B} + 34.1 + \Delta_\mathrm{B,UV}\,, 
\end{equation}
where $L_\mathrm{BOL}$ is the bolometric luminosity of an AGN, $f_B = 10.2$ \citep{1994ApJS...95....1E}, and $\Delta_\mathrm{B,UV} = -0.48$. 

\begin{figure}
\includegraphics[width=\columnwidth]{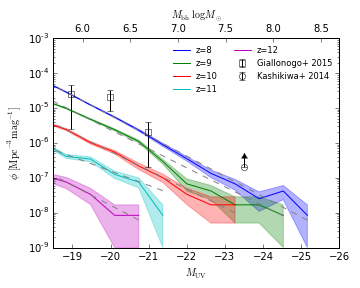}
\caption[AGN UV luminosity function from BlueTides.]
{AGN UV luminosity function from BlueTides. 
  The coloured bands show the $1-\sigma$ sample variance estimated from $100$ sub-volumes (see section \ref{sec:methods}).
  We cut the luminosity function at the faint end at $M_\mathrm{UV} = -18.6$, corresponding to the seed
  mass of the black holes in BlueTides.
  Dashed lines: the best fit power-law model.
  Symbols: measurements at $z=5.75$ by \cite{2015arXiv150202562G, 2015ApJ...798...28K}.}
\label{fig:BHLF}
\end{figure}

Figure~\ref{fig:BHLF} shows the UV luminosity function of AGN in BlueTides.
The luminosity function is cut at $M_\mathrm{UV} = -18.6$, a limit dictated by the imposed seed mass of our black holes ($M_{\rm seed}= 5\times 10^{5} M_{\odot}$). 
The black hole luminosity function is only meaningful for objects that have at least doubled their mass since the black hole was seeded, thereby erasing the artificially imposed seed mass. 
For smaller black holes the AGN luminosity is significantly suppressed due to the artificial absence of black holes in our numerical scheme.
The AGN luminosity function rises steadily at later times, mirroring the evolution in the stellar luminosity function. 
By $z \geq 13$ our box contains a negligible number of AGN and it is thus impossible to reliably estimate the luminosity function.

Unlike the stellar luminosity function, however, the shape of the AGN luminosity function is well-described by a power law.
We thus fit the AGN luminosity function with a power-law model as:
\begin{equation}
\begin{aligned}
\ln \phi(M) &= \ln \phi^\star + \ln A \\ 
    &+ A (M^\star -M)(1 - \alpha_L), 
\end{aligned}
\label{eq:powerlaw}
\end{equation}
where we set a reference magnitude at $M^\star = -18$ without loss of generality.
\citep{1988MNRAS.235..935B,2007ApJ...654..731H} fit the AGN luminosity function at low redshifts with a double power law, which allows a better fit to the steeper bright end slope.
This is not necessary for us as the objects which require such a steeper slope are brighter than any AGN in BlueTides at $z=8.0$.
The best fit parameters of the power law fit are reported in Table \ref{tab:bhfit}.
\begin{table}
  \caption{Best-fit power-Law parameters at $z=8 - 12$ of the AGN UV luminosity function. The fitting model is described by Equation \ref{eq:powerlaw}.}
\label{tab:bhfit}
\centering
\begin{tabular}{ccc}
\hline
$z$ & $\alpha_L$ & $\log \phi^\star$ \\
\hline
8 & $-2.45\pm0.02$ & $-9.33\pm0.04$ \\
9 & $-2.36\pm0.07$ & $-10.44\pm0.09$ \\
10 & $-2.35\pm0.04$ & $-11.91\pm0.07$ \\
11 & $-2.02\pm0.24$ & $-13.79\pm0.33$ \\
12 & $-2.43\pm0.22$ & $-15.30\pm0.28$ \\
\hline
\end{tabular}
\end{table}

\section{Implications For Reionization}
  \label{sec:reion}

There are currently few constraints on the process of hydrogen reionization. Measurements of the total optical depth to the cosmic microwave background (CMB) suggest that the redshift of half reionization is $z_\mathrm{half} \sim 10$ \citep{2013ApJS..208...19H}. Small scale CMB experiments have also constrained the duration of reionization to be $\Delta z < 4.4$. \cite{2012ApJ...756...65Z}.
However, the sources of the UV photons which re-ionized the universe are subject to extensive debate, with the two main candidates being faint galaxies and AGN.
Constraints on the contribution from different sources can be calculated using the measured luminosity functions
\citep[See, e.g.][]{2012MNRAS.425.1413F,2008ApJ...688...85F,2005MNRAS.356..596M,2012ApJ...746..125H,2009MNRAS.394.1812P,
2007MNRAS.382..325B, 2010MNRAS.409..855B, 2013ApJ...768...71R}.

Quasars have yet to be observed at these high redshifts, making the expected impact of AGN on reionization uncertain \citep[see, e.g.][]{2006AJ....132..117F,1993ApJ...412...34M,2012MNRAS.419.1480M}.
However, the most recent constraints on the quasar luminosity function from CANDELS Goods fields at $z\sim 4-6$ suggest that AGN may make a significant contribution to reionization \citep{2015arXiv150202562G}.
In general, reionization driven by rare bright sources such as quasars and large galaxies would progress rapidly, while one driven by faint sources would progress more slowly.

The photon budget can be modelled from first principles using the simulated luminosity functions of ionizing sources. 
As ionizing photons may themselves affect the formation of small halos \citep{2000MNRAS.312L...9M}, direct predictions require simulations which couple radiative transfer and hydrodynamics. 
Furthermore, predicting the UV photon escape fraction from galaxies requires exquisite resolution \citep[see, e.g.][]{2011ASL.....4..228T}.
A simpler approach, which we take here, is to use the simulated luminosity functions from a cosmological hydrodynamic simulation to estimate the photon sources, leaving the escape fraction as a free parameter. 

\begin{figure}
\includegraphics[width=\columnwidth]{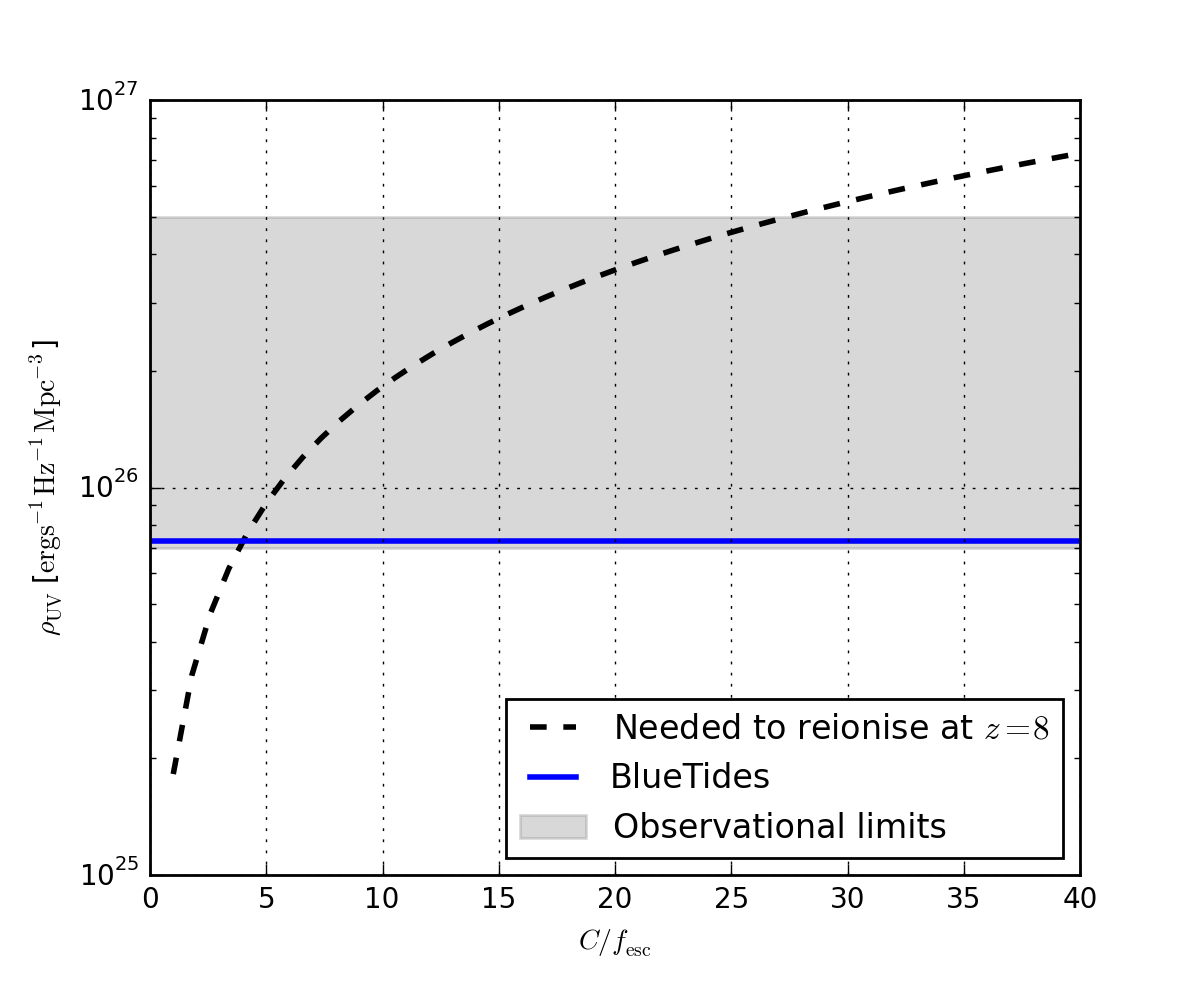}
\caption{Photon budgets to reionise the universe at $z=8$, as a function of the ratio between clumping factor $C$ and escaping fraction $f_\mathrm{esc}$ .
Shaded region: UV emissivity from the observed luminosity function, extrapolated to $M_\mathrm{UV} = -10$.
Black dashed: UV emissivity required for a complete reionization at $z=8$.
Blue solid: UV emissivity from BlueTides, extrapolated to $M_\mathrm{UV} = -10$.
}
\label{fig:reion-8}
\end{figure}

Our modelling of the photo-ionization rate is based on \cite{2012MNRAS.425.1413F}, and we refer the reader to this paper for further details of the model 
\citep[see also][]{2012ApJ...746..125H}. Here we briefly review the relevant pieces. The radiation density $\rho_\nu(z)$ of a source species with a evolving luminosity function $\phi(M_\mathrm{UV}, z)$ and specific luminosity $L_\nu(M_\mathrm{UV}, \nu)$ is
\begin{equation}
\rho_\nu(z) = \int_{M_\mathrm{cut}} 
    \phi(M_\mathrm{UV}, z) L_\nu(M_\mathrm{UV}, \nu) d M_\mathrm{UV}.
\end{equation}
For galaxies, the ionizing photo production, $\Gamma_\mathrm{GAL}$, is based on the star formation rate
\begin{equation}
\Gamma_\mathrm{GAL} (z) = \kappa f_\mathrm{esc}
\frac{\rho^{\mathrm{GAL}}_\mathrm{UV} (z) 
   \unit{M_\odot \mathrm{year}^{-1} \mathrm{Mpc}^{-3}}}
   {1.05\times10^{21} \unit{W \cdot \mathrm{Mpc}^{-3}}},
\end{equation}
where $\kappa=10^{53.1}\,\unit{s^{-1} M_\odot^{-1}\;\mathrm{year} }$ is the mean opacity. $f_\mathrm{esc}$ is the UV escape fraction.
High resolution simulations coupling radiative transfer to hydrodynamics in individual galaxies have suggested a wide range of possible values for $f_\mathrm{esc}$. 
For example, \cite{2008ApJ...672..765G} reported that high redshift galaxies are very inefficient in releasing their photons thus $f_\mathrm{esc} < 5\%$;
\cite{2014ApJ...788..121K} suggests $f_\mathrm{esc} \sim 0.14$.
On the other hand, \cite{2005ApJ...624..485C} favours high escape fractions, while \cite{2011MNRAS.412..411Y} suggests that small halos with halo mass $< 10^9 \msunh$ have $f_\mathrm{esc} \sim 0.4$.
In order to bracket possibilities for the reionization contribution from our galaxies we consider two extreme scenarios: 
a low escape fraction model with $f_\mathrm{esc} = 0.05$, and a high escape fraction model with $f_\mathrm{esc} = 1.0$. 

The AGN contribution, $\Gamma_\mathrm{AGN} (z)$, to the ionizing photon budget depends on the AGN luminosity function and associated SED:
\begin{equation}
\Gamma_\mathrm{AGN} (z) = \int_{\nu_\mathrm{H}}^{\nu_\mathrm{He}} \frac{\rho^\mathrm{AGN}\nu(z)}{h_p \nu} d\nu, 
\end{equation}
where $\nu_\mathrm{H} = 3.2\times10^{15}\unit{Hz}$ and $\nu_\mathrm{He} = 12.8\times10^{15}\unit{Hz}$ are the ionising frequency of Hydorgen and Helium, $h_p$ is the Planck constant, and $\rho^\mathrm{AGN}$ is the radiation density of AGN photons.

\begin{equation}
L_\nu = L_\mathrm{UV} \left(\frac{\nu}{\nu_\mathrm{UV}}\right)^{\alpha_\mathrm{UV}},
\end{equation}
where $\nu_\mathrm{UV} = 2\times10^{15}\unit{Hz}$ is the frequency at $1500$\AA. 
We will adopt $\alpha_\mathrm{UV} = -1.76$ \citep[see, e.g.][]{2007ApJ...654..731H}.

The formulae above provide the photon budget. 
The photo-ionization rate required to reionize the universe can be estimated theoretically from the recombination rate and the clumping factor $C(z)$ as 
\begin{equation}
  \Gamma_\mathrm{REION} (z) = 0.027 \kappa \frac{C}{30} \left(\frac{1+z}{7}\right)^3 \left(\frac{\Omega_b h_{70}^2}{0.465}\right)^2.
\end{equation}
The clumping factor $C(z)$ describes density variations below the resolution of hydrodynamic simulations and can be estimated 
using higher resolution simulations of the intergalactic medium at the reionization epoch \citep{2012MNRAS.427.2464F,2011ApJ...743...82M}.
We use the smallest evolving clumping factor from \cite{2009MNRAS.394.1812P}
\begin{equation}
C(z) = 1 + 43 z^{-1.71}.
\end{equation}
Note that a higher clumping factor requires more photons to reionize the universe.

We extrapolate the luminosity functions measured from our simulations (see Figure \ref{fig:BHLF}) to include the contributions from galaxies and AGN smaller than the resolution limit of the simulation. 
We integrate the galaxy luminosity function for all UV magnitudes brighter than $M_\mathrm{UV} = - 10$, the lower limit of galaxies that generate ionizing photons \citep{2012MNRAS.423..862K}. 
We consider extrapolating the AGN UV luminosity to $M_\mathrm{UV} = -12$ from $M_\mathrm{UV}=-18 $, the faintest AGN in the BlueTides simulation. However, further decreasing the AGN threshold does not increase the number of ionizing photons significantly.

In Figure~\ref{fig:reion-8}, we show the luminosity density in BlueTides at $z=8$ and the photon budget to reionize the universe at $z=8$ as a function of $\frac{C}{f_\mathrm{esc}}$, the ratio between the clumping factor and escaping fraction.
The UV luminosity density in BlueTides, like the luminosity function, is consistent with observations.
We also see that if galaxies alone reionise the universe by $z=8$, BlueTides implies a ratio of $\frac{C}{f_\mathrm{esc}}=4$.  Assuming a clumping factor of $C(z=8) = 2$, reionization at $z=8$ requires a high escape fraction of $f_\mathrm{esc} \sim 60\%$. With less UV photon production, it would be difficult for the galaxies in BlueTides alone to reionize by $z=8$.

\begin{figure*}
\includegraphics[width=1.0\textwidth]{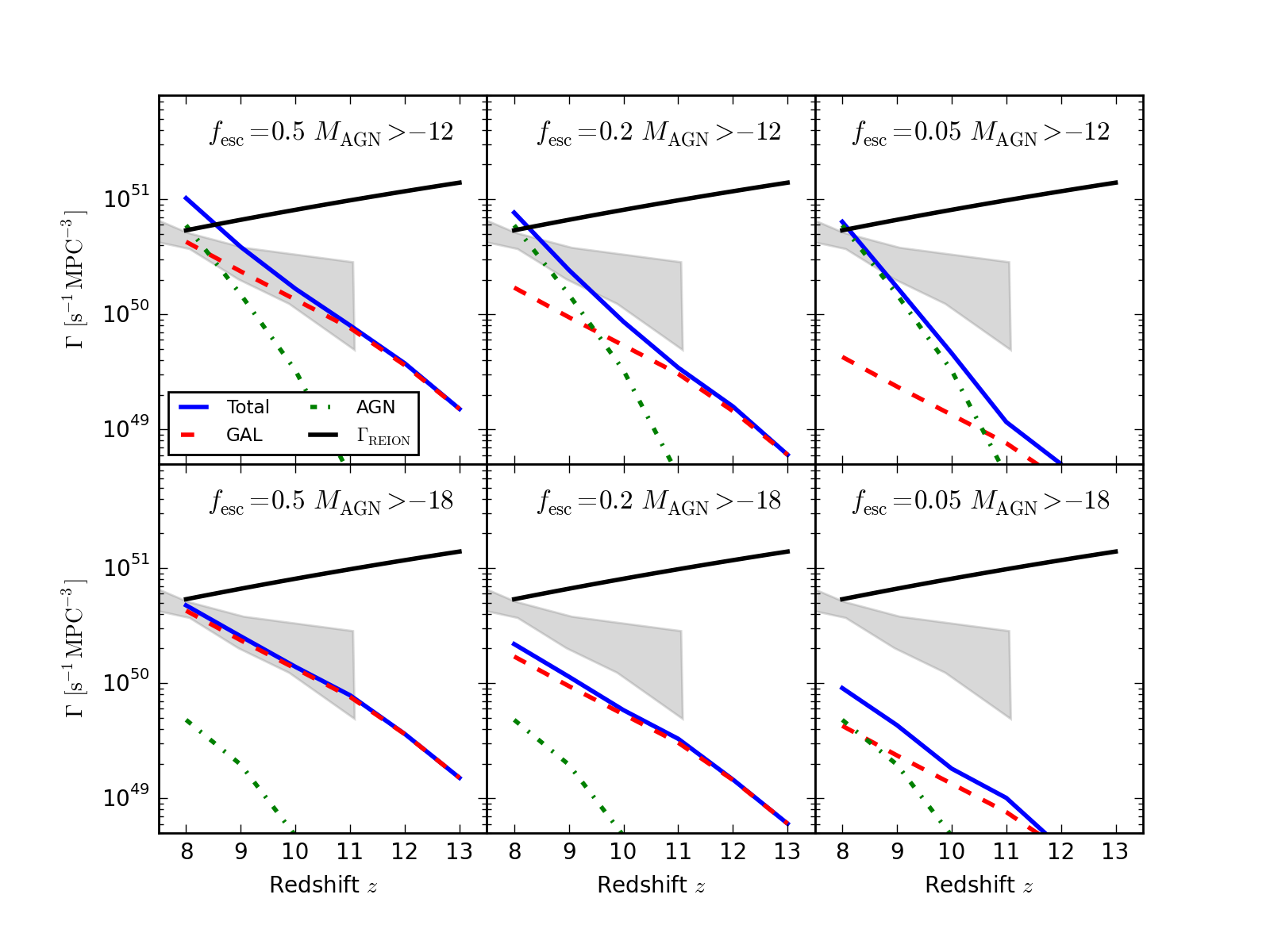}
\caption[Photon budget at $z>8$]{Photon budget at $z>8$.
We consider three models of the escape fraction: 
$f_\mathrm{esc} = 1.0$ (left panels), 
$f_\mathrm{esc} = 0.5$ (middle panels), 
and 
$f_\mathrm{esc} = 0.05$ (right panels).
There are two lower limits for the extrapolated AGN luminosity: 
$M_\mathrm{UV} = -12$ (upper panels), 
and $M_\mathrm{UV} = -18$ (lower panels).
Solid black lines show the theoretical expectation for photons needed to fully ionize the universe.
Solid blue: photo-ionization rate of AGN and galaxies combined.
Dashed red: photo-ionization rate of galaxies.
Dash-dotted green: photo-ionization rate of AGN.
Shaded region: observational constraints combining the optical depth to the CMB, Quasar absorption and Lyman-$\alpha$ emssion, by \citep{2015arXiv150308228B}.
The extrapolation limit of the galaxy luminosity function is fixed at $M_\mathrm{UV} = -10$ (see text).
}
\label{fig:budget}
\end{figure*}

Recent observations (especially Planck) favour reionization at $z<8$. \cite{2015arXiv150308228B} analyzed the constraints on the ionization history, incorporating recent Planck results with constraints from quasar absorption and Lyman-$\alpha$ emission line measurements.
Figure~\ref{fig:budget} shows a comparison between the ionization rate in BlueTides and the observational constraints under various scenarios for the AGN luminosity and galaxy escape fraction.

As mentioned above, provided $f_\mathrm{esc} = 0.5$, galaxies alone can produce an ionization history which completes by $z=8.0$, consistent with current observational constraints.
With a smaller UV escape fraction, the contribution from faint AGN instead drives most of reionization, dominating over the UV photons from galaxies.
As seen in the upper panels of Figure~\ref{fig:budget}, AGN tend to produce an ionization rate which increases faster than that from galaxies. 
It is also interesting to note that the AGN scenario implies that faint AGN exist with luminosities down to  
$M_\mathrm{UV} = -12$, corresponding to a blackhole mass of $10^3\,\mathrm{M_\odot}$ (assuming Eddington accretion), and much smaller than is currently observed. 
Overall, it is difficult to produce reionization completing by $z=8$ without either invoking a high UV escape fraction from galaxies or a significant contribution from a faint AGN population.

\section{Conclusions}
\label{sec:conclusion}
We have performed BlueTides, a high resolution, $400^3\mpch^3$ uniform volume hydrodynamical simulation. 
BlueTides includes a pressure-entropy formulation of smoothed particle hydrodynamics, gas cooling, star 
formation (including molecular hydrogen), black hole growth and models for stellar and AGN feedback processes. 
BlueTides is the first cosmological large volume hydro simulation to incorporate a ``patchy`` reionization model producing an extended hydrogen reionization history.
We have reported the high redshift ($z > 8$) UV luminosity functions of galaxies and AGN in BlueTides, and examined the implications for reionization.

We find good agreement between the expected star formation rate density in BlueTides and current observations at $8 \le z \le 10$.
By using the star formation rate in our galaxies we make predictions for the intrinsic galaxy luminosity functions and show that they compare favourably to observations from Hubble Space Telescope (HST) legacy fields. The brightest galaxies are yet to be observed and we predict that upcoming larger area surveys should start detecting them.
At $z=8$ some dust may be required to reproduce the currently observed bright end in the HST surveys.

Our simulation predicts a faint-end slope of the luminosity function consistent with observations. When fit to 
a Schechter luminosity function, the slope varies between $\alpha \sim -1.8$ at $z=8$ to  $\alpha \sim -2.1$ at $z=10$ with an evolution in the slope $\propto (1+z)^{-0.41}$.
The AGN luminosity functions from BlueTides can be fit by a power law 
with a slope consistent with the most recent observations from CANDELS Goods fields \citep{2015arXiv150202562G}. 
The AGN population evolves quickly at these redshifts with the brightest quasars reaching $M_{\mathrm UV} \sim -25$ at $z\sim 8$, which is 
at least an order of magnitude fainter than SDSS quasars at $z\sim 6$.
By combining the AGN and galaxy luminosity functions we find that the bright end  $M_{\mathrm UV} \sim -21.5$ flattens due to the increased AGN activity above $M_{\mathrm UV} \sim -21.5$.

We find that a high ($\sim 50\%$) escape fraction is still required for galaxies alone to produce enough photons to reionize the Universe by $z=8$. 
Our high escape fraction model supports the conditions proposed by \cite{2012MNRAS.423..862K}: a reionization model that includes mostly galaxies
requires both an extrapolation to very faint-end $M_\mathrm{UV} = -10$ and a sharp increase of the escape fraction (up to $50\%$) at high redshift, in agreement with \cite{2015arXiv150308228B}.
For lower escape fractions (closer to $10\%$-$20\%$), a possible source of the extra photons are faint AGN with black hole masses as faint as $M \sim 10^3 M_\odot$ \citep{2004ApJ...604..484M}.
\cite{2015arXiv150202562G} has suggested that the faint end of the AGN luminosity function at $z < 5.75$ may favour these sources. AGN would lead to a relatively quick reionization which could soon be testable observationally. Alternatively, reionization may not complete until $z < 8$, as suggested by \cite{2014arXiv1410.5439F}. and consistent with the recent optical depth measurements from Planck \citep{2015arXiv150201589P}. 

\section*{Acknowledgement}
We acknowledge funding from NSF OCI-0749212
and NSF AST-1009781. The BlueTides simulation was run on 
facilities at the National Center for Supercomputing Applications. 
The authors thank Dr. Nishikanta Khandai for discussions at the planning stage of the simulation; and NCSA staff members for their help in accommodating the run at BlueWaters.

\bibliographystyle{mn2e}
\bibliography{mybib}

\end{document}